\documentclass{moriond}
\usepackage{graphicx}
\usepackage{wrapfig}
\usepackage{subcaption}
\usepackage{amsmath}
\usepackage{float}
\usepackage{url}
\usepackage[capitalise]{cleveref}

\def\be{\begin{equation}}
\def\ee{\end{equation}}
\def\bea{\begin{eqnarray}}
\def\eea{\end{eqnarray}}

\newcommand{\Photo}{}

\begin{document}
\vspace*{4cm}
\title{Status and Prospects of PADME}

\author{E. Long on behalf of the PADME Collaboration\footnote{S. Bertelli, F. Bossi, R. De Sangro, C. Di Giulio, E. Di Meco, D. Domenici, G. Finocchiaro, L.G. Foggetta, M. Garattini, A. Ghigo, P. Gianotti, M. Mancini, I. Sarra, T. Spadaro, E. Spiriti, C. Taruggi, E. Vilucchi, (INFN Laboratori Nazionali di Frascati), V. Kozhuharov (Faculty of Physics, University of Sofia ``St. Kl. Ohridski'', and INFN Laboratori Nazionali di Frascati), K. Dimitrova, S. Ivanov, Sv. Ivanov, R. Simeonov (Faculty of Physics, University of Sofia ``St. Kl. Ohridski''), G. Georgiev (Faculty of Physics, University of Sofia ``St. Kl. Ohridski'' and INRNE, Bulgarian Academy of Science), F. Ferrarotto, E. Leonardi, P. Valente, A. Variola (INFN Roma1), E. Long, G.C. Organtini, M. Raggi (Physics Department, ``Sapienza'' Universit\`a di Roma and INFN Roma1), A. Frankenthal (Department of Physics, Princeton University)}}

\address{Dipartimento di Fisica, Sapienza Universit\`a di Roma, Rome, RM, Italy}

\maketitle\abstracts{
The Positron Annihilation to Dark Matter Experiment (PADME) was designed and constructed to search for dark photons ($A'$) in the process $e^+e^-\rightarrow\gamma A'$, using the positron beam at the Beam Test Facility (BTF) at the National Laboratories of Frascati (LNF). Since the observation of an anomalous spectra in internal pair creation decays of nuclei seen by the collaboration at the ATOMKI institute, the PADME detector has been modified and a new data-taking run has been undertaken to probe the existance of the so-called ``X17" particle.}

\section{Introduction}\label{sec:Introduction}

\begin{figure}[H]
	\begin{subfigure}{0.5\linewidth}
		\includegraphics[scale=0.61]{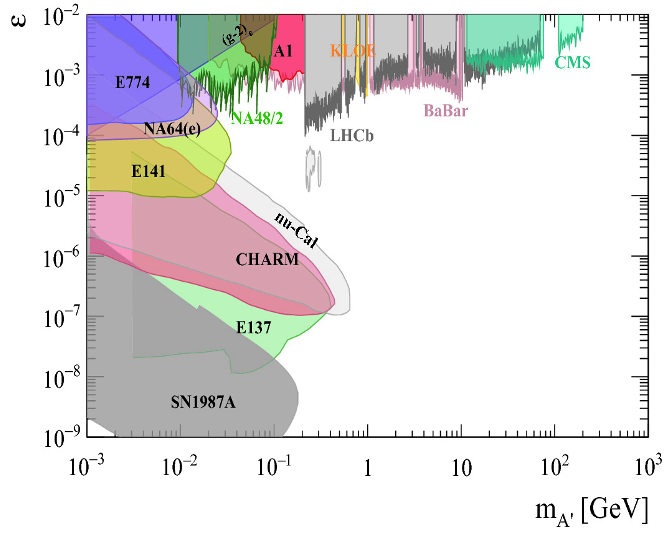} 
       \caption{Constraints for visibly decaying $A'$s.}
       \label{subfig:VisibleSearches}
        \centering
	\end{subfigure}%
%	\hfill
	\begin{subfigure}{0.5\linewidth}
		\includegraphics[scale=0.61]{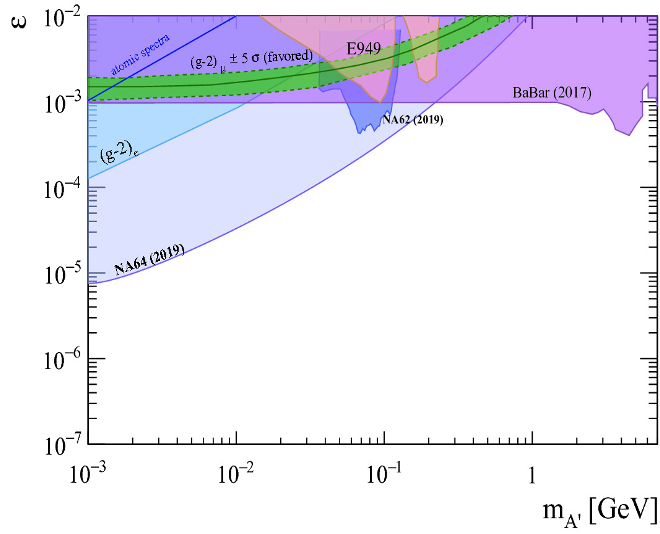} 
       \caption{Constraints for invisibly decaying $A'$s.}
       \label{subfig:InvisibleSearches}
        \centering
	\end{subfigure}
	\caption{Current constraints on Dark Photon models \protect\cite{Fabbrichesi:2020wbt}. Models are characterised by two variables: the mass, $m_{A'}$, and the coupling, $\varepsilon$, between the $A'$ and SM fermions.}
	\label{fig:DPConstraints}
\end{figure}

As the parameter space available to WIMP models of dark matter has reduced over recent years, interest has grown in dark sector models. These models assume that dark matter is made of particles which interact feebly with Standard Model particles via a portal particle, known as a dark photon ($A'$). The dark photon would be a massive vector boson, characterised by two parameters: the mass ($m_{A'}$) and the coupling ($\epsilon$) to Standard Model fermions. Current constraints for Dark Photon models are shown in \Cref{subfig:VisibleSearches,subfig:InvisibleSearches} respectively \cite{Fabbrichesi:2020wbt}.

The Positron Annihilation to Dark Matter Experiment (PADME) was constructed to search for a dark photon ($A'$) produced in association with a standard model photon in $e^+e^-\rightarrow\gamma A'$ annihilations. Since switching on in 2018 the experiment has collected more than $10^{13}$ positrons on target (PoT) and has published the most precise measurement of the $e^+e^-\rightarrow\gamma\gamma$ cross section below 1~GeV to date \cite{PADME:2022tqr}.

\begin{wrapfigure}[30]{L}{0.45\textwidth}
	\centering
	\includegraphics[scale=0.6]{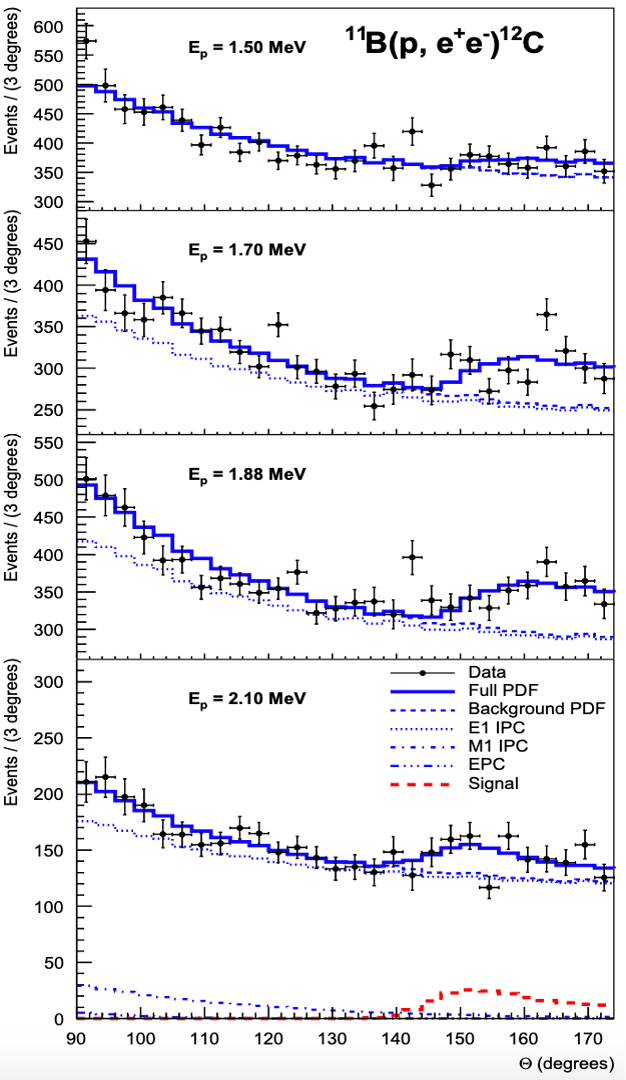} 
    \caption{$^{12}$C decays for four different excitation energies \protect\cite{Krasznahorkay:2022pxs}.}
    \label{fig:ATOMKIAngularSpec}
\end{wrapfigure}

This talk gave an update on the status and prospects of PADME Run 3 which aims to investigate the anomaly found in internal pair creation (IPC) decays of different nuclei by Krasznahorkay et al. at the ATOMKI Institue in Debrecen, Hungary. This anomaly, also known as the ``X17 anomaly", was first found in the angular spectrum of $e^+e^-$ pairs resulting from $^8$Be decays. The collaboration found a 6.8$\sigma$-significance bump in the spectrum at separation angles corresponding to the decay of a particle of approximately 17~MeV mass \cite{Krasznahorkay:2015iga}. The experiment has since been repeated with an improved setup \cite{Krasznahorkay:2019lyl}, using the decays of $^4$He and $^{12}$C. The collaboration has found that the anomaly persists in the decays of these nuclei too, with three different excitation states being studied in $^4$He \cite{PhysRevC.104.044003} and four states of $^{12}$C \cite{Krasznahorkay:2022pxs}. All three of the $^4$He results gave a significance above 6$\sigma$ for a particle of approximately 17~MeV mass decaying to $e^+e^-$ pairs, and the $^{12}$ results having significances 3$\sigma$, 3$\sigma$, 7$\sigma$ and 8$\sigma$ respectively. \Cref{fig:ATOMKIAngularSpec} shows the results from the decays of $^{12}$C, and \Cref{fig:X17Constraints} shows the current exclusion limits on the vector boson and ALP interpretations of the X17 anomaly respectively, along with the expected 90\% confidence limit reachable from PADME Run 3 \cite{Darme:2022zfw}.

%\begin{figure}[H]
%	\begin{subfigure}{0.48\linewidth}
%	  \centering
%		\includegraphics[scale=0.53]{Images/He4Original.png} 
 %      \caption{$^4$He decays for three different excitation energies \cite{PhysRevC.104.044003}.}
 %      \label{subfig:He4Original}
%	\end{subfigure}%
%	\hfill
%	\begin{subfigure}{0.48\linewidth}
%	  \centering
%		\includegraphics[scale=0.53]{Images/C12Original.png} 
  %     \caption{$^{12}$C decays for four different excitation energies \cite{Krasznahorkay:2022pxs}.}
 %      \label{subfig:C12Original}
%	\end{subfigure}
%	\caption{Opening angle spectrum of $e^+e^-$ from IPC decays.}
%	\label{fig:ATOMKIAngularSpec}
%\end{figure}

PADME is located at the Beam Test Facility (BTF) at the National Laboratories of Frascati (LNF), the only facility in the world with a positron beam able to operate at energies of $\sim$280~MeV%, the resonance energy of a 17~MeV particle
. Therefore, in 2022 the PADME collaboration decided to search for the resonant production and subsequent decay of a new 17~MeV boson.

\section{Setup and status of PADME: Runs 1 and 2}\label{sec:SetupRuns1and2}
\begin{figure}[b]
	 \centering
	 \includegraphics[scale=0.25]{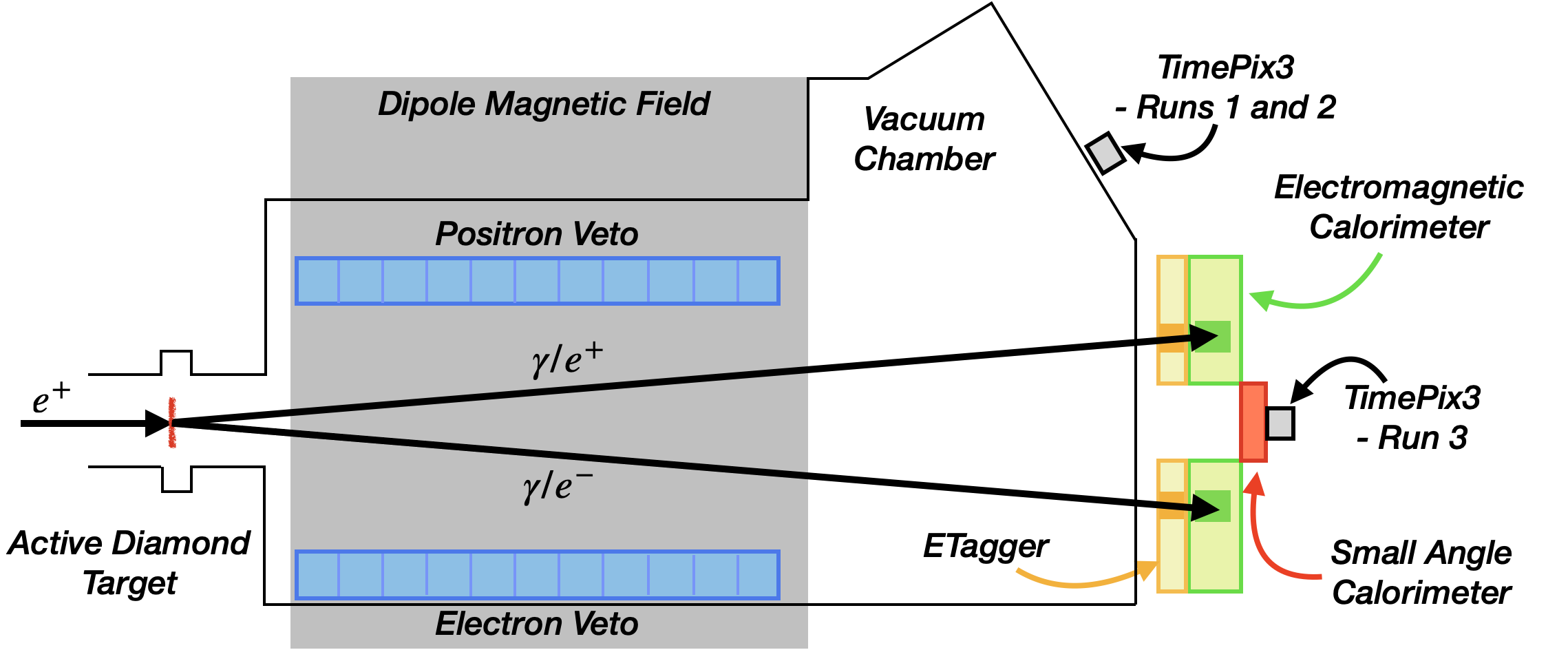}
	 \caption{PADME Setup}
	 \label{fig:MixedSetup}
 \end{figure}

PADME is a fixed target experiment, with the $<550$~MeV positron beam impinging on the 100$\mu$m thin active diamond target. The target acts as a luminosity detector, measuring the number and geometrical distribution of positrons in each beam bunch, as well as being the target for the physical processes studied at PADME. The dark photon signal which the experiment was designed to search for is a single standard model photon in the BGO electromagnetic calorimeter (ECal) and nothing in the other detectors. The missing energy then gives access to the mass of the dark photon ($m_{A'}$). The experimental setup can be seen in \Cref{fig:MixedSetup}. It should be noted that the Charged Particle Tagger (ETagger) was added for Run 3, whereas the Dipole Magnetic Field was used only in Runs 1 and 2 and switched off for Run 3. The position of the TimePix3 beam monitor was changed accordingly, since un-interacted beam particles were deflected by the magnetic field in Runs 1 and 2 and not in Run 3.

The main background at PADME is Bremsstrahlung. The rate of Bremsstrahlung photons is so high that it would overwhelm the ECal if no measures were taken, however since the photon spectrum is sharply peaked at small angles to the beamline, the ECal was built with a central hole, behind which a faster Cherenkov detector known as the small angle calorimeter (SAC) was constructed. In Runs 1 and 2 the magnetic field diverted charged particles leaving the target into the charged particle veto system, allowing Bremsstrahlung photons to be matched with their accompanying positrons, giving a more efficient veto system. The pair of charged particle vetoes also gives access to visible decay modes of the dark photon in Runs 1 and 2. More information about the PADME detector can be found in \cite{PADMEComissioning}.

\begin{wrapfigure}[17]{R}{0.5\textwidth}
	 \centering
	 \includegraphics[scale=0.95]{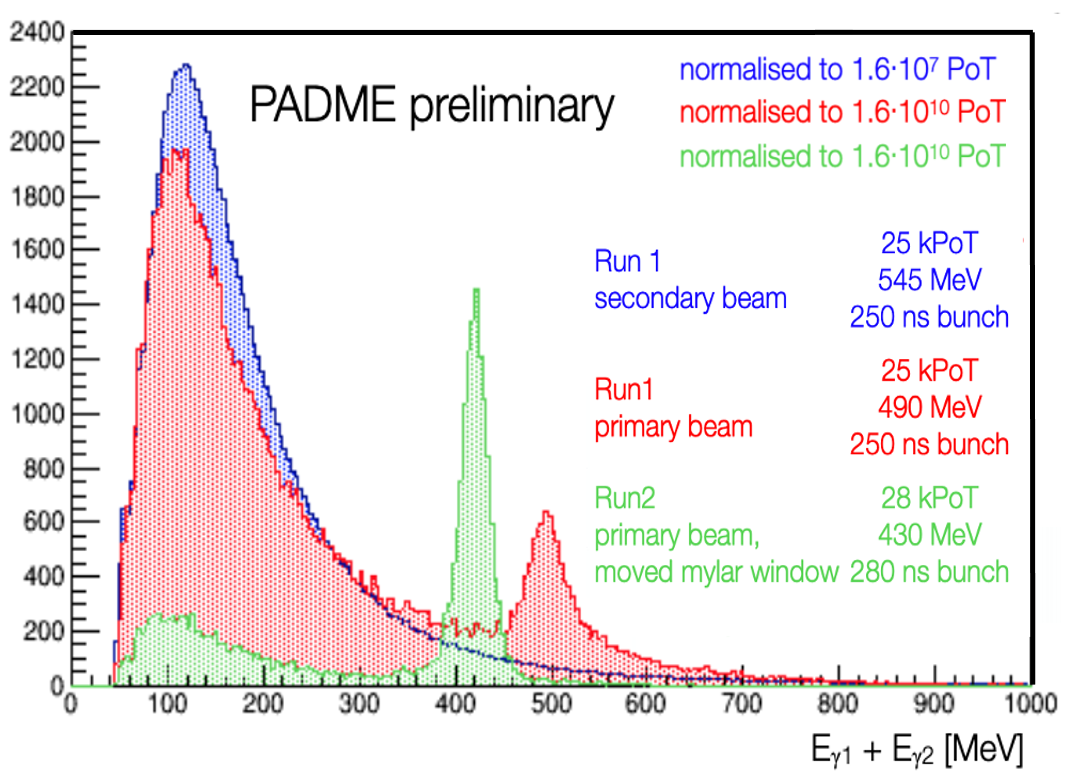}
	 \caption{Sum of energies of photon pairs in PADME ECal in three beamline configurations between Runs 1 and 2.}
	 \label{fig:PhotonPairEnergyRuns1and2}
 \end{wrapfigure}

Between 2018 and 2020, the PADME collaboration undertook 2 data-taking runs, collecting a total of $>10^{13}$PoT in three different beam-line configurations. As shown by the total energy of pairs of photons reaching the ECal, \Cref{fig:PhotonPairEnergyRuns1and2}, the background in the ECal coming from low energy beam particles was reduced by moving from the secondary positron beam, produced at a 1.7X$_0$ Cu target close to the PADME experimental hall (shown in blue), to the primary positron beam produced at the LINAC positron converter further upstream (shown in red). The situation was improved significantly more by moving the vacuum separating window upstream of a wall separating the LINAC and BTF, and changing the material from which it was made from beryllium to mylar (shown in green). This last modification was made after careful studies using a Monte Carlo simulation of the beamline and the experiment \cite{PADME:2022ysa}.

\section{Multi-photon annihilation}\label{sec:gammagamma}
The beam conditions shown in green in \Cref{fig:PhotonPairEnergyRuns1and2} allowed the collaboration to measure the inclusive in-flight cross-section $e^+e^-\rightarrow\gamma\gamma(\gamma)$. This measurement represents an important milestone for the collaboration for several reasons: first, it could be sensitive itself to new physics at the Sub-GeV scale, for example in the form of ALPs. It is also the first measurement of this cross-section at beam energies below 500~MeV with precision better than 20\%, and the only measurement of this cross-section in this range which observed the two final state photons, making it the first measurement of this cross-section able to distinguish SM processes from new physics contributions. Finally, it is an important milestone in the search for the $A'$ associated production, since the cross section of associated production of a Dark Photon is linearly correlated to the cross section of 2$\gamma$ $e^+e^-$ annihilation ($\sigma(e^+e^-\rightarrow\gamma\gamma)$) as:
$$\sigma(e^+e^-\rightarrow\gamma A')\propto \varepsilon^2\times\sigma(e^+e^-\rightarrow\gamma\gamma)\times\delta(m_{A'})$$
where $\varepsilon$ is the coupling constant of the $A'$ to SM leptons and $\delta(m_{A'})$ is an acceptance factor that is a function of the mass of the Dark Photon.

\begin{wrapfigure}[17]{L}{0.5\textwidth}
    \centering
	 \includegraphics[scale=0.25]{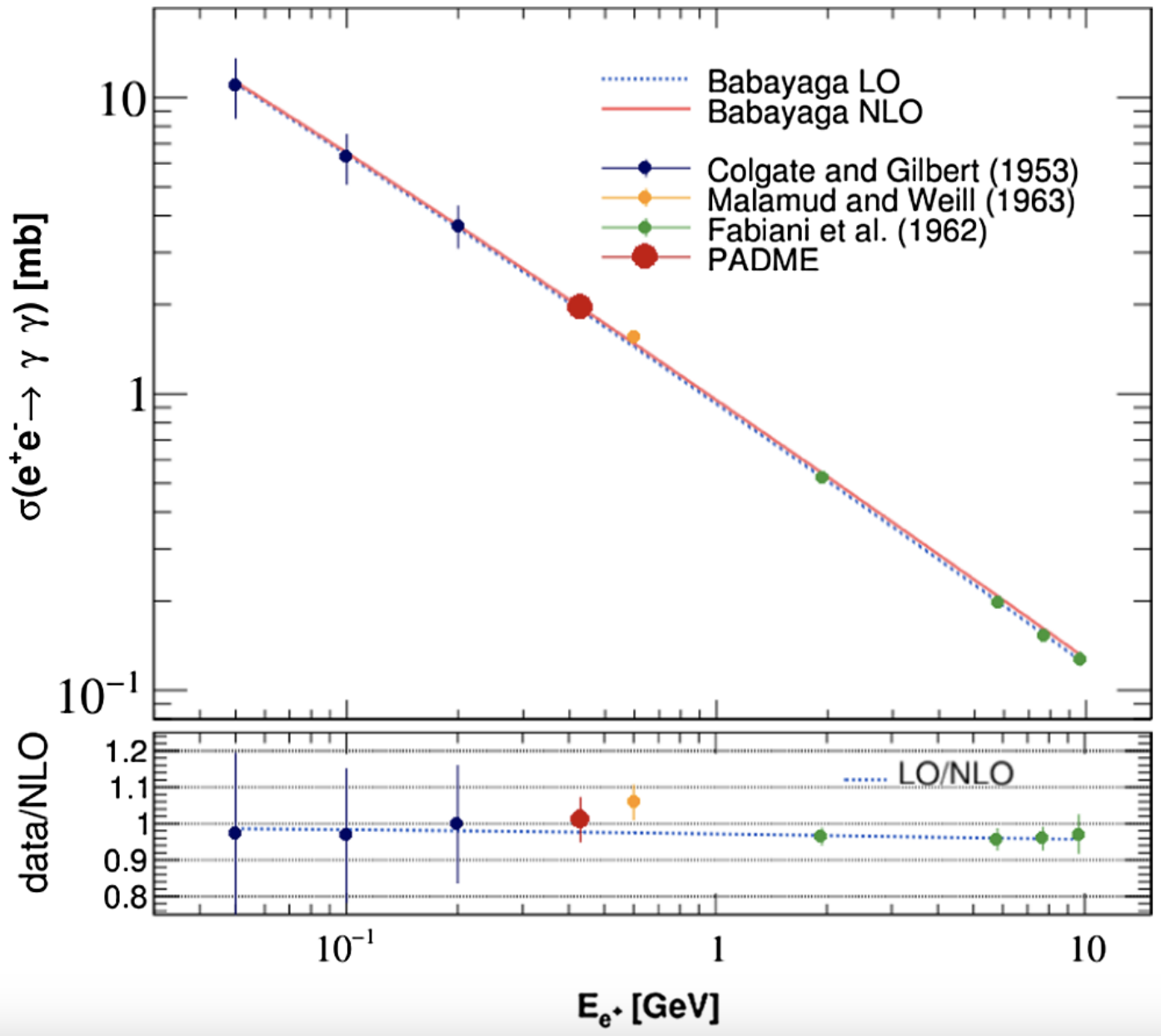}
	 \caption{$\sigma(e^+e^-\rightarrow\gamma\gamma)$ measurements at the $<$10~GeV scale. The measurement from PADME Run 2 is shown in red.}
	 \label{fig:GammaGammaCrossSec}
\end{wrapfigure}

Using the tag-and-probe method presented in \cite{PADME:2022tqr}, the cross-section was found to be 
$$\sigma(e^+e^-\rightarrow\gamma\gamma)_{PADME} = 1.977 \pm 0.018\ (stat) \pm 0.119\ (syst)\ \text{mb}$$
which agrees with QED at next to leading order, as calculated by Babayaga, 
$$\sigma(e^+e^-\rightarrow\gamma\gamma)_{Theory} = 1.9478 \pm 0.0005\ (stat) \pm 0.0020\ (syst)\ \text{mb}$$
within the experimental error. The result is shown in the context of other measurements at this energy scale in \Cref{fig:GammaGammaCrossSec}.

%\begin{figure}[H]
%\end{figure}
 
\section{PADME Run 3}\label{sec:PADMERun3}
\begin{figure}[]
	\begin{subfigure}{0.5\linewidth}
		\includegraphics[scale=0.24]{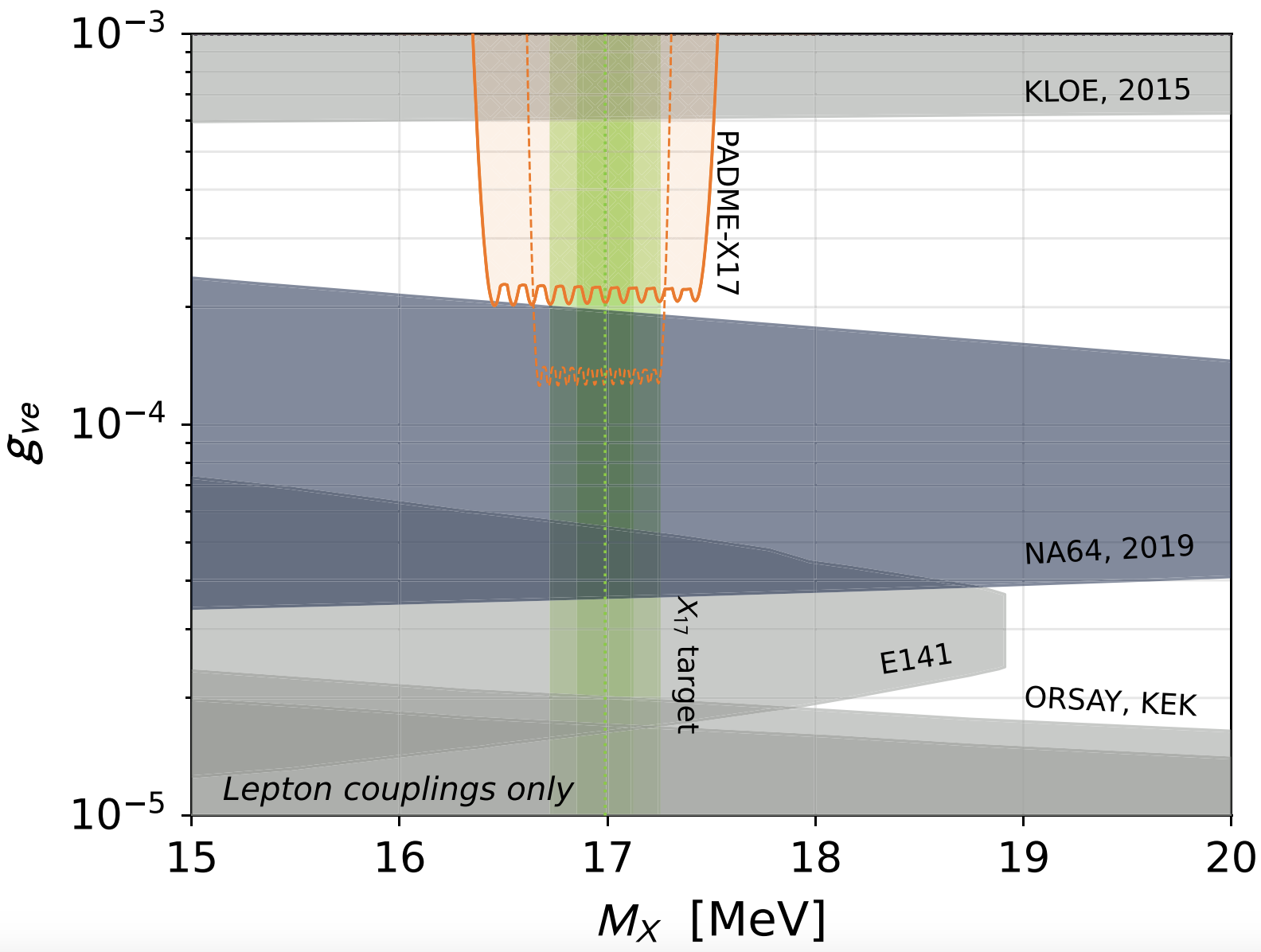} 
       \caption{Constraints for vector X17 interpretation.}
       \label{subfig:X17VectorExclusion}
        \centering
	\end{subfigure}%
%	\hfill
	\begin{subfigure}{0.5\linewidth}
		\includegraphics[scale=0.24]{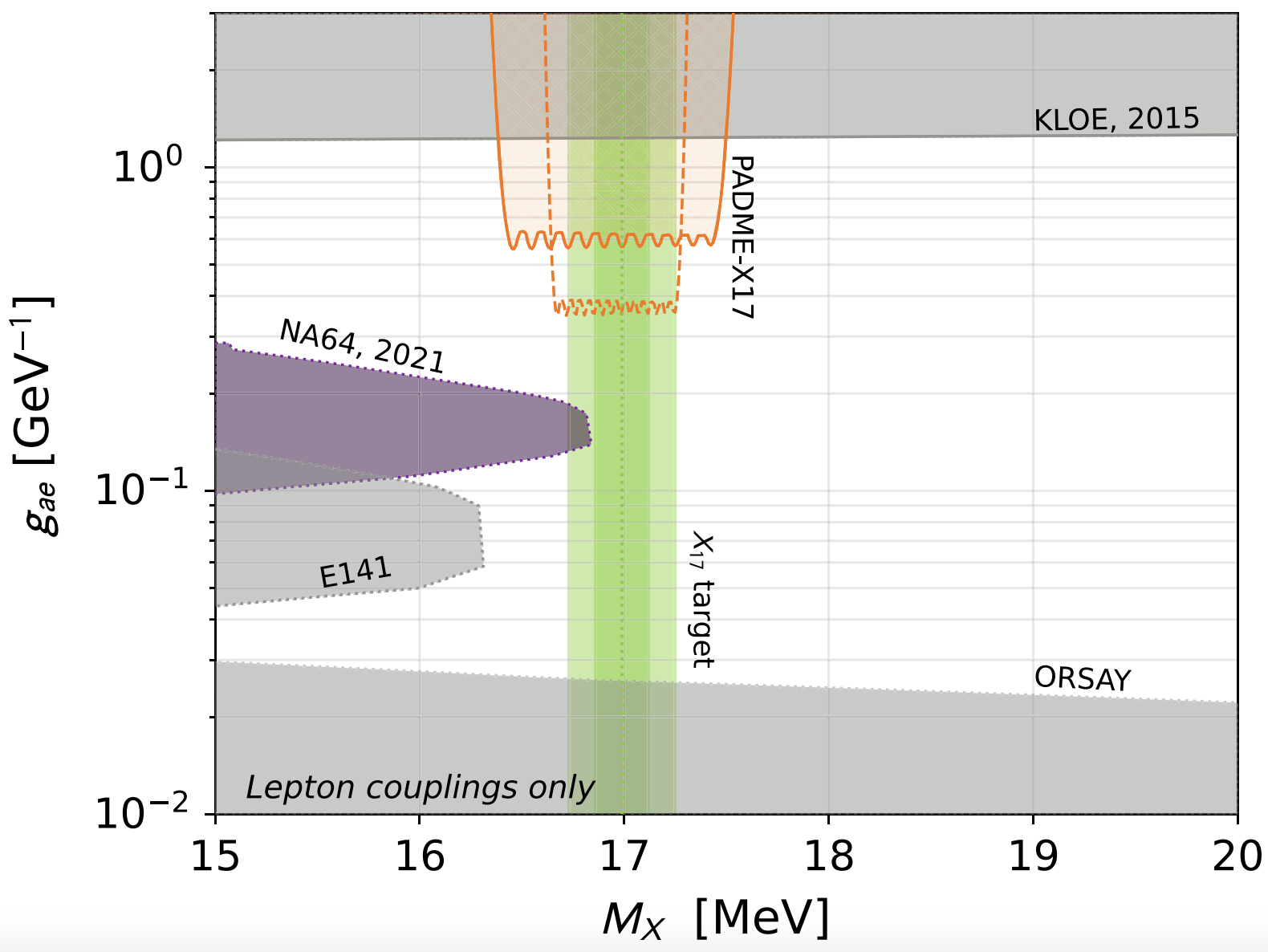} 
       \caption{Constraints for ALP X17 interpretation.}
       \label{subfig:X17ALPExclusion}
        \centering
	\end{subfigure}
	\caption{Current constraints on X17 models \protect\cite{Darme:2022zfw}. The targeted parameter space for PADME Run 3 is shown in orange.}
	\label{fig:X17Constraints}
    \end{figure}

As discussed in \Cref{sec:Introduction}, in 2022 the PADME collaboration undertook a new round of data-taking specifically searching for the X17 particle. Due to the expected increase in the cross section of production of the particle on resonance, the collaboration decided to perform a scan across the energy range expected from the ATOMKI experiments, shown as the green band in \Cref{fig:X17Constraints}. Studies from Run 2 revealed the difficulty of studying charged-particle final states using the Vetoes. For this reason, the experimental setup was modified in order to perform the measurement using the ECal: the PADME dipole was switched off and a new plastic scintillator detector, known as the ETagger was built directly in front of the ECal, as shown in \Cref{fig:MixedSetup}.
 
A scan was performed over beam energies between 260-300~MeV in steps of 0.7~MeV, with the beam energy being selected by changing the current of the penultimate LINAC dipole magnet before the PADME target, and the beam trajectory was then corrected using the last dipole along the beamline in order to redirect the beam back along the PADME axis. While in Runs 1 and 2 the beam multiplicity was 28$\times$10$^3$ PoT per bunch, with the magnetic field off this would cause the ECal to be overwhelmed with Bremsstrahlung positrons. Therefore, the multiplicity was reduced to 5$\times$10$^3$ PoT per bunch by keeping the LINAC collimators relatively closed. This had the benefit of allowing in turn for a very low energy spread at each point, corresponding to an excellent invariant mass resolution. Due to the expected increase in production cross-section on resonance, only $10^{10}$ PoT were required for each point in the scan, corresponding to approximately 24 hours of data-taking even at this lowered multiplicty.

\begin{figure}[]
	\begin{subfigure}{0.5\linewidth}
		\includegraphics[scale=0.3]{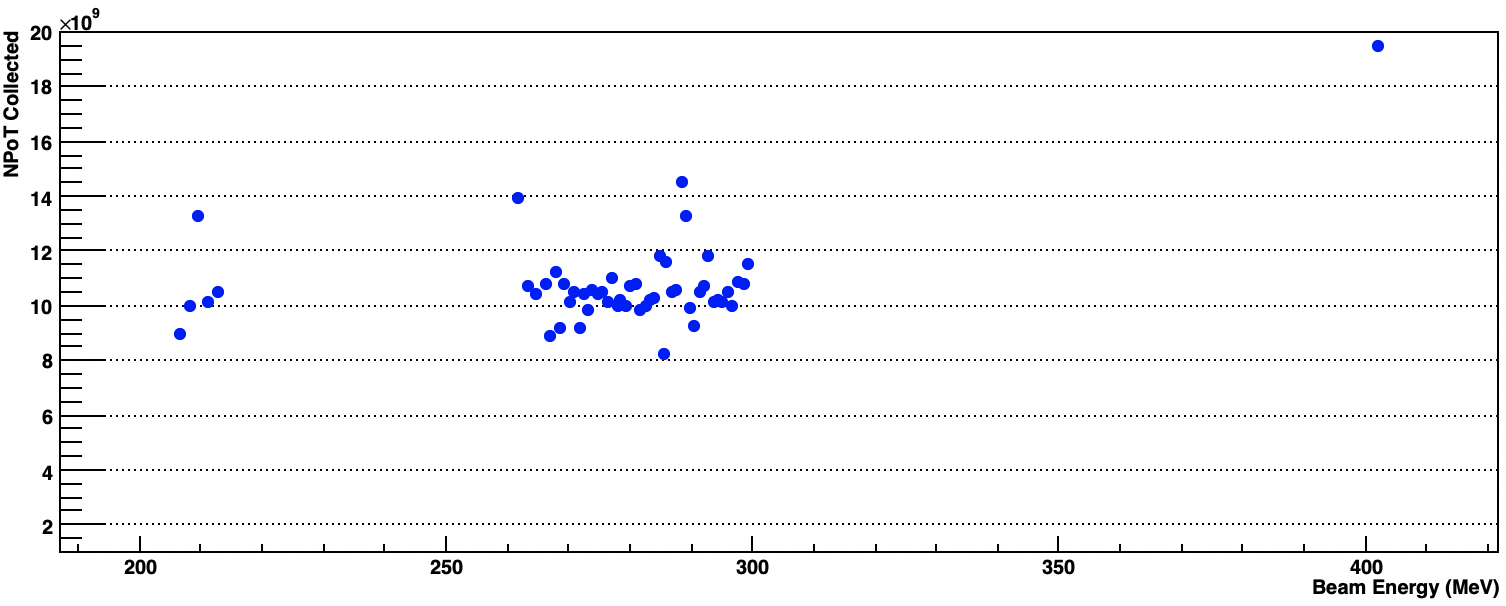} 
       \caption{PoT as a function of beam energy.}
       \label{subfig:NPoTEBeam}
        \centering
	\end{subfigure}%
%	\hfill
	\begin{subfigure}{0.5\linewidth}
		\includegraphics[scale=0.3]{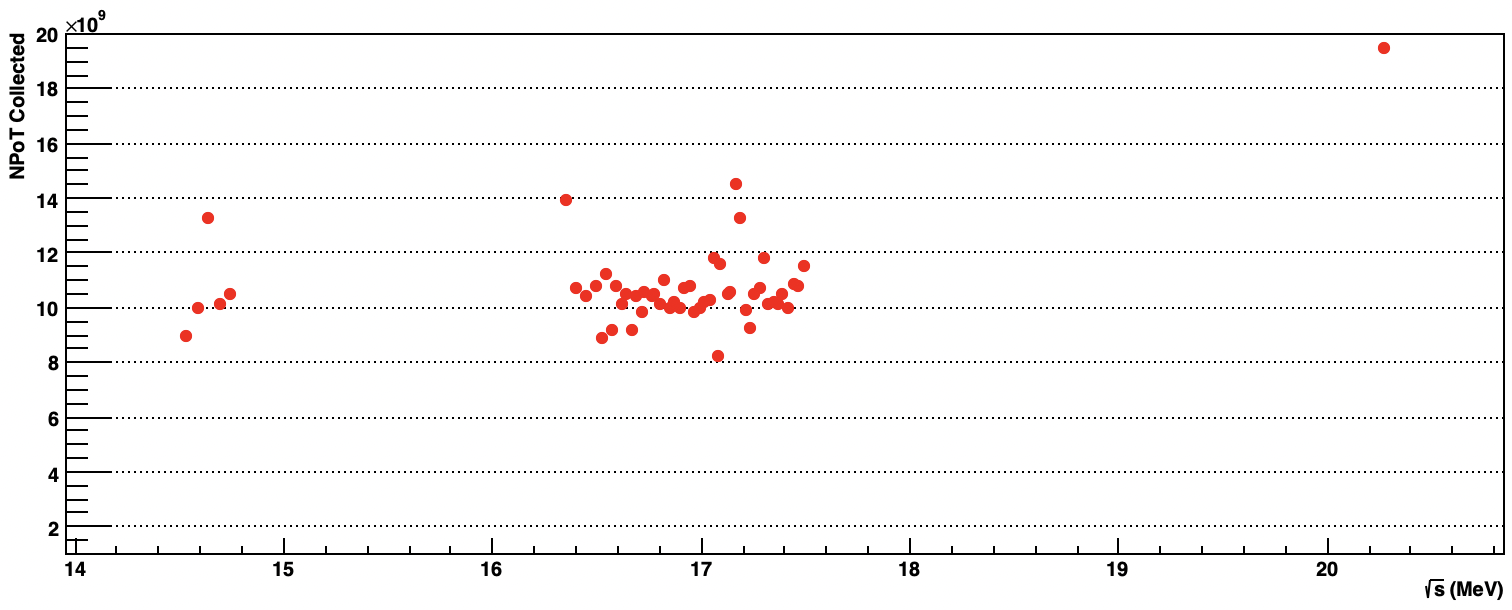} 
       \caption{PoT as a function of center of mass energy.}
       \label{subfig:NPoTSqrtS}
        \centering
	\end{subfigure}
	\caption{Number of positrons on target collected in Run 3 as a function of beam energy and center of mass energy respectively.}
	\label{fig:NPoT}
\end{figure}

$10^{10}$ PoT were collected at each of 47 points around the X17 resonance, 5 points below the resonance and 1 point above the resonance, as shown in \Cref{fig:NPoT}. The off-resonance points serve to validate the analysis strategy before unblinding the signal region. The X17 signal would then appear as a bump above the SM background in the number of events in several bins around one point of the scan. The expected limit at PADME, given the statistics accumulated shown in \Cref{fig:NPoT}, is of the same density as the lower orange lines in \Cref{fig:X17Constraints}, with the range of the upper orange lines.

%\vspace{1cm}
%\vspace{1cm}

\begin{figure}[H] 
        \begin{minipage}[t][6cm]{0.49\textwidth}
		\includegraphics[scale=0.76]{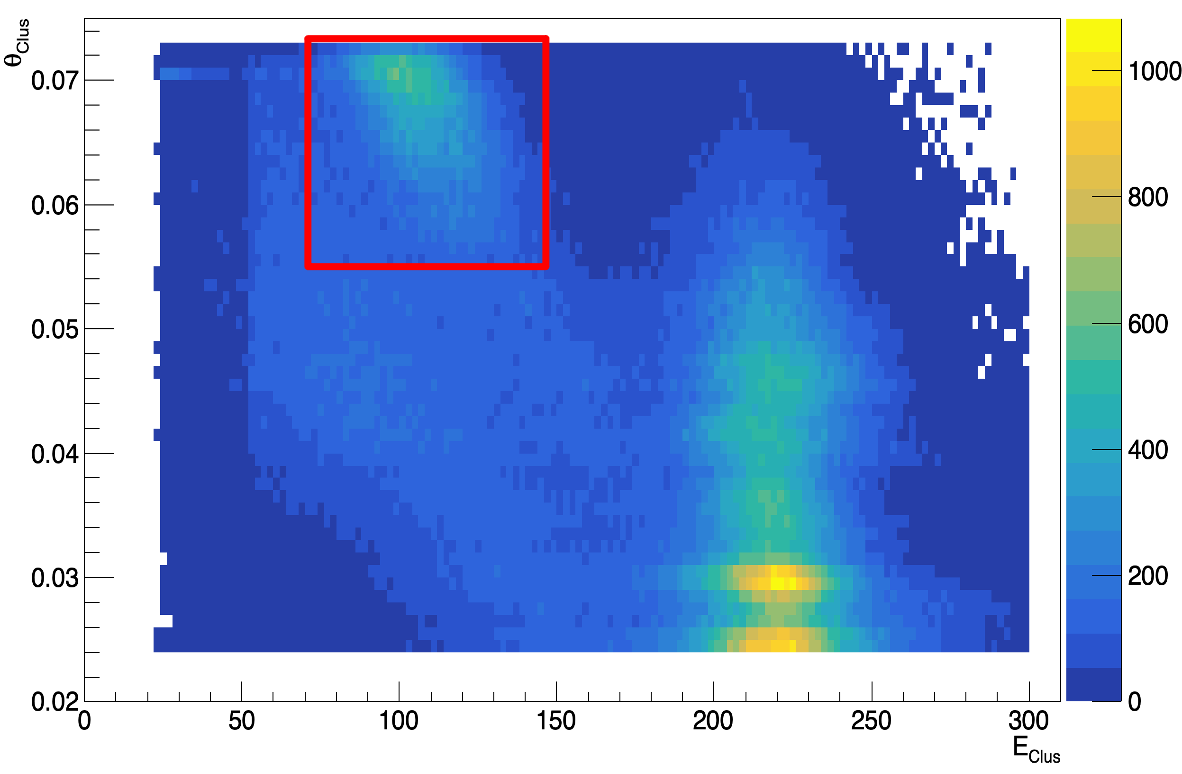} 
            \caption{Energy of clusters in ECal vs theta angle to beamline.}
            \label{fig:EVsThetaECal}
            \centering
	\end{minipage}    
        \hfill
	\begin{minipage}[t][6cm]{0.49\textwidth}
	    \includegraphics[scale=0.17]{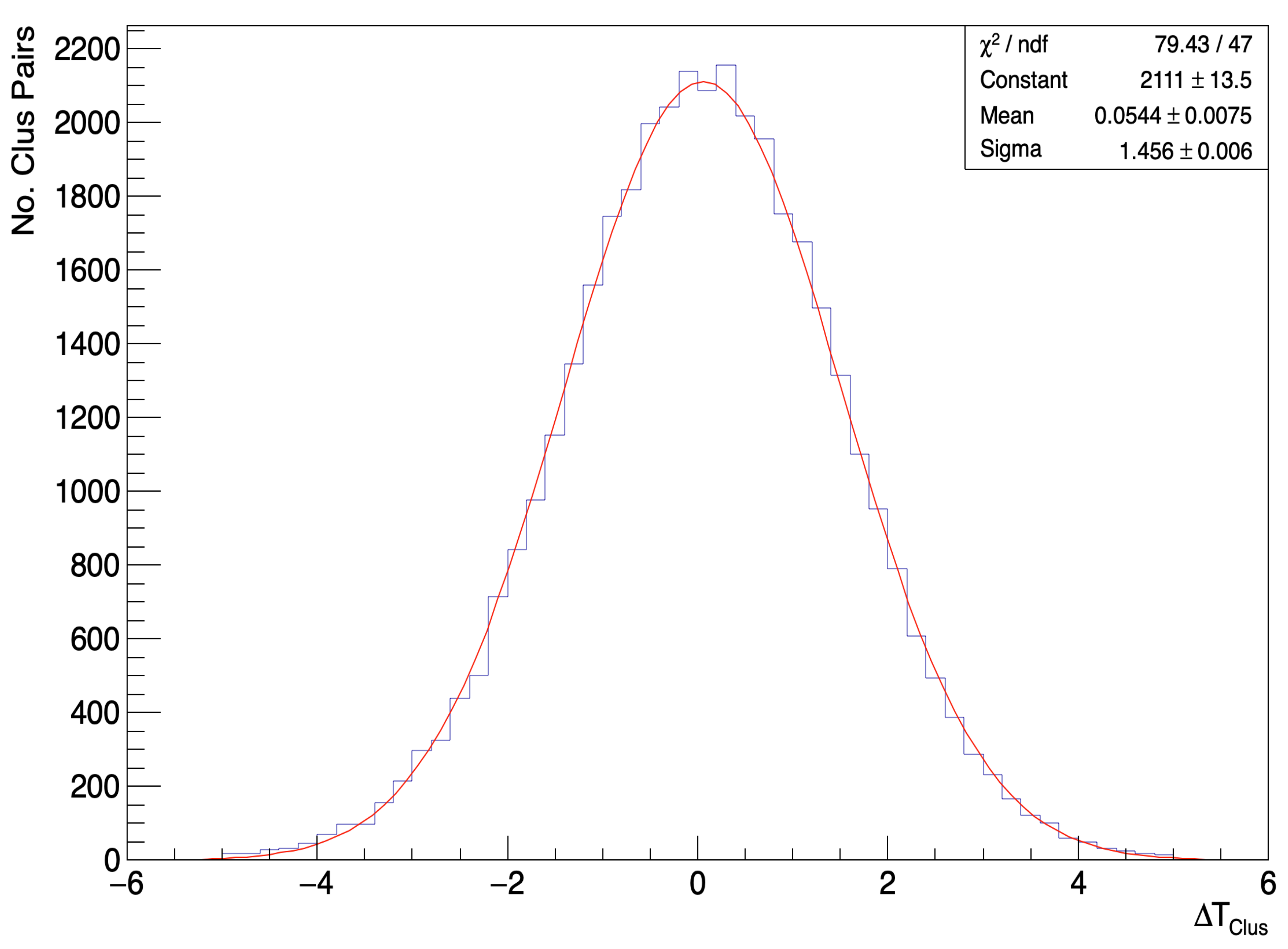} 
            \caption{Time difference of cluster pairs within red box in \protect\Cref{fig:EVsThetaECal}.}
            \label{fig:DeltaTECalClus}
            \centering
        \end{minipage}
    \end{figure}

As of April 2023, the data has been inspected to assess the data quality. \Cref{fig:EVsThetaECal} shows the energy of all clusters in the ECal as a function of the theta angle between the cluster and the beamline, for all 5 of the below resonance points. Due to kinematic constraints, any particles coming from vertices at the target should have a kinematic profile which lies inside the box highlighted in red 
%between 55~mrad and the edge of the ECal acceptance at 0.73~mrad and between 70~MeV and 140~MeV,
. Studying the difference between time of arrival of two clusters in this box gives \Cref{fig:DeltaTECalClus}. The fact that this distribution is very well fitted with a gaussian of $\sigma\sim 1.5$~ns shows that the experiment is not dominated by background either from out-of-trajectory beam particles or from pile-up, and that the data is in a good condition to be analysed.

\section{Conclusion}
PADME was designed and constructed to search for dark photons in $e^+e^-$ annihilation. In 2022 the collaboration published its first physics measurement, finding the cross-section $e^+e^-\rightarrow\gamma\gamma$ to be well in agreement with the Standard Model at next to leading order. Later that year the collaboration turned its attentions to the X17 anomaly found at the ATOMKI institute, undertaking a specific data-taking run searching for this particle on-resonance. Inspections of the data quality show that the data is not dominated by backgrounds coming from the beamline, and that therefore it is in a good state for the X17 analysis.

\section*{Acknowledgments}
This work has been mainly funded by Istituto Nazionale di Fisica Nucleare. Other funds have been also granted as part of MUCCA, CHIST-ERA-19-XAI-009, and TA-LNF as part of STRONG-2020 EU Grant Agreement 824093 projects.

\section*{References}

\bibliography{Bibliography.bib}

\end{document}